\documentclass[aps,pre,twocolumn,superscriptaddress,showpacs,showkeys ]{revtex4-1}
\usepackage{color}
\usepackage{graphicx}
\usepackage{hyperref}
\usepackage{amsmath,amssymb}
\usepackage{epstopdf}
\usepackage{subcaption}
\graphicspath{{figs/}}
\usepackage{float}
\usepackage[singlelinecheck=false,justification=raggedright]{caption}
\usepackage{dcolumn}   % needed for some tables
\usepackage{bm}        % for math
\usepackage{verbatim}   % for math

\begin{document}
\title{A Self-Organized Model for the Flicker Noise in Interacting Two-Dimensional Electron Gas}

\author{Maryam Pirgholi}
%\email{valizadeh.neda1204@gmail.com}
\affiliation{Department of Physics, University of Mohaghegh Ardabili, P.O. Box 179, Ardabil, Iran}

\author{Morteza Nattagh Najafi}
\email{morteza.nattagh@gmail.com}
\affiliation{Department of Physics, University of Mohaghegh Ardabili, P.O. Box 179, Ardabil, Iran}

\author{Vadood Adami}
%\email{valizadeh.neda1204@gmail.com}
\affiliation{Department of Physics, University of Mohaghegh Ardabili, P.O. Box 179, Ardabil, Iran}

\begin{abstract}
We investigate self-organized criticality in a two-dimensional electron gas (2DEG) by introducing a lattice-based model that incorporates electron-electron interactions through the concept of coherence length. Our numerical simulations demonstrate that in the strongly interacting regime, the system exhibits a distinct set of universal critical exponents, markedly different from those observed in the weakly interacting limit. This dichotomy aligns with experimental findings on the metal-insulator transition in 2DEGs, where high interaction strength (low carrier density) leads to qualitatively different behavior. The analysis includes scaling of the average electron density with temperature, the power spectral density, and the statistics of electronic avalanches—namely their size distributions and autocorrelation functions. In all cases, the extracted exponents differ significantly between the weak and strong interaction regimes, highlighting the emergence of two universality classes governed by interaction strength. These results underscore the critical role of electron correlations in the self-organized behavior of low-dimensional electronic systems.
\end{abstract}

%\pacs{05., 05.20.-y, 05.10.Ln, 05.45.Df}
%\keywords{{\color{blue}Depinning Transition, porous media}, fluid dynamics, critical exponents}

\maketitle

\section{Introduction}
The transport characteristics of parabolic systems involving two-dimensional semiconductors have been a persistent issue in condensed matter physics. A key question in this domain has been whether a two-dimensional (2D) electronic system can exhibit a metallic ground state, a topic that has sparked considerable debate over an extended period~\cite{abrahams2001metallic,kravchenko2003metal}. 
According to the one-parameter scaling theory of quantum localization developed by Abrahams and colleagues, a true metallic phase is forbidden in two-dimensional systems when spin interactions and spin-dependent scattering are absent~\cite{abrahams1979scaling}.
Although some experimental evidence supports this prediction, such as the logarithmic increase in resistance as \(T \to 0\) observed in silicon MOSFETs~ \cite{dolan1979nonmetallic}, counterexamples have surfaced. Notably, Kravchenko's findings~\cite{kravchenko1995scaling}, challenge this prediction within the strong interaction regime. Kravchenko et al. demonstrated that in samples with minimal disorder and in regimes of strong electron interactions (\(r_s \gg 1\)), a metallic phase unexpectedly emerges, where the resistivity (\(\rho\)) increases with temperature (\(\frac{d\rho}{dT} > 0\)). In ref~\cite{najafi2019electronic} the authors designed a dynamical percolation theory for the observed metal-insulator transition in 2D.\\

Flicker noise in two-dimensional electron gases (2DEG) represents another enigmatic phenomenon in dense low-dimensional systems, with its origin still not fully understood. In natural systems, flicker noise is associated with signals whose power spectrum exhibits a power-law decay and heavy-tailed behavior at low frequencies, typically following the form \( \frac{1}{f^\alpha} \), $\alpha$ known as the flicker exponent. The observed values of the exponent \(\alpha\) typically fall within the range \([1.1 - 1.5]\)~ \cite{kravchenko2003metal, weissman19881}. Since the exponent \(\alpha\) is typically close to 1, flicker noise is often referred to as \( \frac{1}{f} \) noise~\cite{zhang20001}. These systems exhibit oscillations predominantly at very low frequencies.
Besides the electrical systems ~\cite{carreras2004evidence}, one may mention solar flares~\cite{bogdan2001avalanche}, stock market price fluctuations~\cite{matia2003multifractal}, rainfall ~\cite{dhar1990self}, and earthquakes~\cite{bak1989earthquakes} as some examples. This behavior is attributed to the absence of characteristic scales in the system’s dynamics. Numerous experiments have reported the observation of \( \frac{1}{f} \) noise in various quantities, such as the electronic resistance \( R(t) \) of a 2DEG, measured under fixed temperature conditions and specific perturbations. Instances of \( \frac{1}{f} \) noise have also been identified in vacuum tubes~ \cite{johnson1925schottky}, as well as in carbon composites and thick-film resistors~\cite{barry2014measurement}. \\

Interestingly, flicker noise emerges spontaneously without the need for external parameter tuning, a hallmark of self-organized critical (SOC) systems—an extensive class of out-of-equilibrium systems characterized by intrinsic scale separation~\cite{dhar2006theoretical}. If this framework applies to 2DEG, a fundamental question arises: what self-organizing mechanism underlies the emergence of this phenomenon within the 2DEG system?
The quasi-equilibrium state of such SOC systems—where ``quasi" reflects the near-equilibrium nature of the electronic transitions—plays a crucial role in advancing the understanding of flicker noise in 2DEGs. Numerous studies have attempted to explain the observed \( \frac{1}{f} \) noise in these systems; however, most efforts have been largely phenomenological and have seldom succeeded in fully capturing the underlying mechanisms of the phenomenon.\\

Many natural processes (\cite{halley1996ecology,beggs2003neuronal,bertotti2006science,field1995superconducting,petri1994experimental,salminen2002acoustic,chang2003complexity,chang1999self}) can be described by simple local rules and display characteristics of SOC~ \cite{bak1989earthquakes,dhar1990self}, with most showing power-law decay behavior in the power spectrum of their temporal signals imposing strong constraints on their spatial and temporal correlations~\cite{dhar1990self}. Some promissing results came out asserting that BTW sandpile model (as a prototypical example for SOC systems without need for fine-tuning of external parameters~\cite{bak1988self}) can describe the flicker noise~\cite{maslov19991,chang2003complexity,chang1999self,jensen19891}. Numerous efforts have been made to identify the structure of temporal and spatial correlations and anti-correlations in sandpiles ~\cite{hwa1992avalanches,kutnjak1996temporal,ali1995self,lubeck2000crossover}, with further discussions available in ~\cite{font2015perils,janicevic2018threshold}.  It was asserted that long-term correlations between avalanches in SOC systems can lead to the emergence of \( \frac{1}{f^\alpha} \) noise \cite{davidsen20021}. It was then shown that the flicker noise is absent in the BTW sandpile model~\cite{najafi2021flicker} (equivalently, $\alpha=2$ which trivially describes systems where the autocorrelation function decays exponentially~\cite{boffetta1999power}).
\\

Here in this paper we examine the model proposed in~\cite{najafi2019electronic} designed for 2DEG to inspect if it shows Flicker noise. This model addresses key deficiencies in the existing literature by providing a unified framework that captures the role of electron-electron interactions in driving self-organized critical behavior in 2DEGs. Unlike previous studies that often neglect interaction effects or treat them perturbatively, the approach employed in this model incorporates interactions non-perturbatively through a coherence-length-based discretization. This allows us to identify two distinct universality classes corresponding to weak and strong interaction regimes, offering a coherent explanation for the experimentally observed differences in scaling behavior across the metal-insulator transition.\\

The structure of the paper is organized as follows: In the next section, we present the theoretical framework of the proposed model along with the discretization scheme based on the coherence length. Section III provides the results of our numerical simulations, accompanied by a detailed analysis of the critical exponents in both weakly and strongly interacting regimes. Finally, the concluding section summarizes our findings, discusses their theoretical implications, and outlines possible directions for future research.

\section{The problem definition and motivation}
The primary types of noise encountered in solid state physics include Nyquist noise, shot noise, and flicker (or pink) noise. The power spectrum associated with thermal noise in a resistor with resistance \( R \) at temperature \( T \) is constant in terms of frequency $f$ described by the relation \( PS_v(f) = 4k_B T R \), where \( k_B \) is the Boltzmann constant, while the power spectrum of shot noise, which is discrete due to the transfer of charge carriers, depends on the quantity of charge and the magnitude of the current \cite{raychaudhuri2002measurement}. In contrast, the flicker noise, also referred to as pink noise or \( \frac{1}{f} \) noise, is another significant form of noise, where the power spectrum decays in a power-law fashion in terms of frequency, that will be introduced in the following section, and will be thoroughly investigated for 2DEG in this project.

\subsection{Flicker Noise in 2DEG}
The designation ``flicker" arises from its distinctive behavior, characterized by rapid fluctuations in noise amplitude over time. This type of noise is characterized by an unconventional power spectrum exhibiting a \( \frac{1}{f^\alpha} \) dependence where $\alpha$ is close to one in 2DEG. It is typically observed in systems carrying current at low frequencies, as well as various out-of-equilibrium  systems, like the  natural systems, neural networks and economic contexts~\cite{musha1992noise,voss19921}. In condensed matter systems, the fundamental origin of this behavior remains under active discussion, while a widely supported hypothesis attributes this phenomenon to the tunneling of electrons within a disordered background \cite{johnson1925schottky}.\\ 

The manifestation of this phenomenon in 2DEG is particularly intriguing, given its significance in the study of Anderson and Mott transitions as functions of temperature and disorder strength \cite{byczuk2005mott}. In 2DEG, flicker noise has been widely observed, with its origins linked to several factors, including material disorder~\cite{dutta1981low} and electron-electron interaction~\cite{dutta1981low}. The primary objective of this project is to explore the behavior of flicker noise in a 2DEG and to construct its phase diagram as a function of interaction strength and temperature, for a fixed disorder intensity.\\

\begin{figure}\centering
\includegraphics[width=8cm, height=6cm]{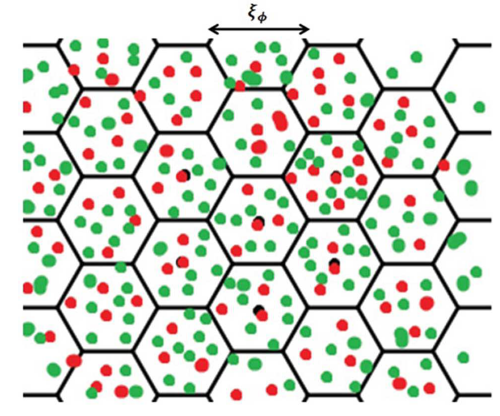}   	      \caption{ Two-dimensional electron gas with hexagonal meshing. Green (red) particles represent electrons (impurities).}
\label{Two-dimensional electron}
\end{figure}

A well-accepted strategy for explaining the flicker noise in the electronic systems is through using the percolation theory~\cite{kogan2008electronic}. It provides a mathematical framework for studying the properties of randomly connected networks. In the context of 2DEG, percolation theory has been employed to explain the emergence of charge transport behavior and to account for conductivity fluctuations in disordered materials~\cite{najafi2018percolation}. In this strategy, a charged carrier moves in the background of electronic puddles characterized by random charges~\cite{li2011disorder}.\\
Our model consists of dynamical puddles formed by the dynamical background carrier density, and also includes a threshold in the chemical potential above which electrons transition into neighboring regions. Like the SOC systems~\cite{dhar2006theoretical}, our model includes also the concept of separation of time scales arisen from the threshold dynamics explained above. The characteristic flickering behavior is attributed to electron emission governed by avalanche dynamics. By introducing the concept of coherence length ($l_{\phi}$), which characterizes phase decoherence due to disorder in a 2DEG, we can distinguish two distinct regimes of electron dynamics: those occurring at length scales much smaller than $l_{\phi}$ and those at scales much larger. In the former regime, a quantum mechanical description is employed, whereas in the latter, a semi-classical transport model is utilized~\cite{backes2015observation}. Following this approach, the system is modeled using hexagonal blocks, as illustrated in Fig.~\ref{Two-dimensional electron}, where the energy associated with each block is given by the one for a coherent quantum electron gas.

\subsection{A review on the model}

In this section, we elaborate on the theoretical model introduced in~\cite{najafi2021flicker}. As briefly discussed in the preceding section, the central concept underlying this model is the discretization of the 2DEG into spatially distinct regions, referred to as coherent cells. Each cell possesses a characteristic linear dimension denoted by $l_\phi$, corresponding to the local phase coherence length. These regions are hypothesized to emerge intrinsically due to disorder and impurity-induced scattering.\\
Within each coherent cell, the 2DEG is treated as a locally coherent quantum electronic system. Intercellular charge transport is facilitated through quantum tunneling processes, governed by the differences in free energy between adjacent cells. \\
The internal energy and chemical potential for each coherent cell (assumed to exhibit a uniform electron density) are computed using the Thomas-Fermi-Dirac (TFD) approximation. The average energy of the $i$-th cell is given by~\cite{najafi2021flicker,najafi2019electronic}:
\begin{equation}
\langle E_i \rangle = K(T,\tilde N_i) + V_{ee}(T, \tilde N_i) + E_{\text{imp}}(T,\tilde N_i)
\end{equation}
where $T$ is the absolute temperature ($\beta = \frac{1}{k_BT}$, $k_B$ is the Boltzmann constant in the following arguments), $\tilde N_i$ denotes the number of electrons in cell $i$, $K(T,\tilde N_i)$ is the thermal-averaged kinetic energy, $V_{ee}(T,\tilde N_i)$ is the electron-electron interaction energy, and $E_{\text{imp}}(T, \tilde N_i)$ is the energy contribution from  the impurities. Using the TFD and the Gaussian local density approximation~\cite{najafi2021flicker,najafi2019electronic}
the total energy is found to be:
\begin{equation}
\begin{split}
E(\left\lbrace N \right\rbrace,&\left\lbrace Z \right\rbrace ,T) =\sum_{i=1}^N\left\langle E_i \right\rangle\\
&=\sum_{i=1}^{L} \left[ -\alpha T^2  \, \mathrm{Li}_2\left(1 - e^{\frac{N_i}{T}}\right) + \beta_0 N_i^2 - \gamma_i N_i \right],
\end{split}
\label{Eq:EnergyMain}
\end{equation}
where $\mathrm{Li}_2(x)$ denotes the polylogarithm function of order two and
\begin{equation}
\begin{split}
&\alpha = 2m \left( \frac{\pi k_B l_\phi(T)}{\hbar} \right)^2,\ \beta_0 = \frac{1}{8\sqrt{2} \varepsilon_0 l_\phi(T)} \left( \frac{e\alpha}{k_B} \right)^2, \\
&\gamma_i = \frac{\alpha e^2\sinh^{-1}(1)}{\pi \varepsilon_0 k_B l_\phi(T)} Z_i,\ N_i = \frac{k_B}{\alpha}\tilde{N} _i.
\end{split}
\end{equation}
In this equation $e$ ($m$) is electron charge (mass), $\varepsilon_0$ is the vacuum permittivity, and $Z_i$ represents the local disorder parameter in the $i$-th cell, sampled from a uniform distribution defined as~\cite{najafi2021flicker}:
\begin{equation}
P(Z) = \frac{1}{\Delta} \, \Theta\left( \frac{\Delta}{2} + (Z - Z_0) \right) \Theta\left( \frac{\Delta}{2} - (Z - Z_0) \right),
\end{equation}
where $\Delta$ is the disorder strength, $Z_0$ is a mean value, and $\Theta(x)$ is the Heaviside step function. Additionally, we suppose that
\begin{equation}
\langle Z \rangle = Z_0, \quad \langle Z_i Z_j \rangle = \delta_{ij}
\end{equation}
where $\delta_{ij}$ is the Kronecker delta and $\langle \ \rangle$ indicates the ensemble average. To generate impurities, for each lattice site $i$, a random number $r \in [0, 1]$ is drawn from a uniform distribution, and the corresponding value of $Z_i$ is computed using the appropriate mapping relation, ensuring that the disorder is sampled according to the prescribed probability distribution:
\begin{equation}
Z_i = Z_0 + \Delta (r - 0.5).
\end{equation} The coherence length $l_\phi(T)$ behaves in a power-law fashion with temperature, and in 2D it is beleived that $l_\phi(T) = a T^{-1/2}$, with $a$ being a constant. 

\begin{figure}[H]
\centering
\includegraphics[width=5cm, height=5cm]{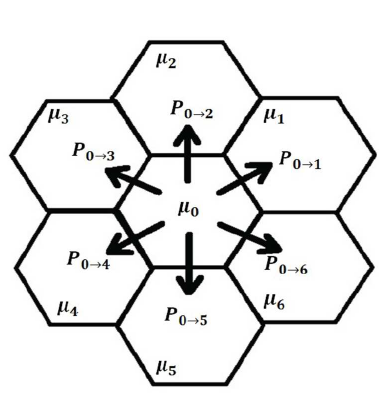}      
\caption{Schematic representation of the electron transition to neighboring cells, where $\mu_0$ represents the donor cell.}
\label{The schematic}
\end{figure}

To extract the chemical potential we use the identity~\cite{najafi2021flicker}:
\begin{equation}
A_{\tilde{N}}(V,T) - T \left( \frac{\partial A_{\tilde{N}}}{\partial T} \right)_{\tilde{N},V} = \langle E \rangle
\end{equation}
where \( A_{\tilde{N}}(V, T) \) denotes the Helmholtz free energy of the cell at volume \( V \) and temperature \( T \). This equation leads to the following relation for the chemical potential $\mu_i\equiv \frac{\partial A_i}{\partial \tilde{N}}$:
\begin{equation}
\mu_i = k_B T \ln \left( e^{h_i} - 1 \right) + U T^{1/2} h_i - I T^{1/2} Z_i,
\end{equation}
where $h_i = \frac{N_i}{T}$, and the coefficients $U$ and $I$ are defined as:
\begin{align}
U &= \frac{2k_B m \sqrt{2Da} e^2 \pi^2}{8 \varepsilon_0 \hslash^2},\ I = \sinh^{-1}(1) \frac{e^2}{\pi \varepsilon_0 \sqrt{Da}}
\end{align}
with $D$ and $a$ representing system-dependent constants. Finally, the chemical potential difference between two neighboring cells 1 and 2 is obtained as~\cite{najafi2021flicker}:
\begin{equation}
\begin{split}
\mu_2 - \mu_1 = k_B T \ln\left( \frac{e^{h_2} - 1}{e^{h_1} - 1} \right) + U T^{1/2}(h_2 - h_1) -\\ I T^{1/2}(Z_2-Z_1).
\end{split}
\end{equation}
We use this formula in the charge transport between cells. We use also a similar equation for \textit{injecting} electrons to the system. More precisely, since the system is coupled to an electronic reservoir the electrons are allowed to be injected to or extracted  from the system, provided that its local chemical potential satisfies special conditions. Upon the injection of an electron, the local electron density is redistributed such that electrons tend to move from regions of higher local chemical potential to neighboring regions with lower local chemical potential (as schematically illustrated in Fig.~\ref{The schematic}).\\

In accordance with Ref.~\cite{najafi2021flicker} we consider a triangular lattice, and the system initially begins with a random configuration. An electron is randomly injected at a site $i_0$ from an external source (electron reservoir). This site is labeled as \textit{unstable} if its chemical potential content exceeds a global chemical potential $\mu_0$ given externally to the system, i.e., $\mu(i_0) > \mu_0$. For such unstable sites the possibility of transferring the electrons to the neighboring sites is examined and executed if permissible. This process may render the adjacent sites unstable, releasing electrons toward their own neighbors, ultimately leading to a cascade of charge transfers throughout the system. The dynamics continues until all sites reach a locally stable configuration. This chain may potentially initiate an electron avalanche of size $S$, defined as the total electronic \textit{topplings} occured in an avalanche. After the system reaches stability, an equilibration process begins, during which random sites are selected to test local charge redistribution. Upon completion of this phase, a new site is randomly chosen for electron reinjection. A probabilisic approach is employed to govern the charge transfer, according to which an electronic injection to site $i_0$ occures with the probability:
\begin{equation}
\begin{split}
P_{\text{reservoir}\to i_0} & \propto  \exp\left[-\beta\left( \mu_{i_0}-\mu_0\right)\right].
\end{split}
\end{equation}
The probability for an electron to hop from cell $i$ to a neighboring cell $j$ follows the Boltzmann factor:
\begin{equation}
    P_{i\to j}\propto \exp\left[-\beta (\mu_j - \mu_i)\right].\\
\end{equation}

To model this transport mechanism, we adopt the Metropolis Monte Carlo method, which serves as a semi-classical approximation to the quantum dynamics of tunneling. In this formalism, the spatial distribution of electrons is governed by the internal energy—dependent on both temperature and the local chemical potential~\ref{Eq:EnergyMain}. For instance, if the donor cell has potential $\mu(0)$ and its neighbors have potentials $\mu(j)$ for $j = 1, 2, \dots, 6$, and assuming $\mu(i) < \mu(j)$ for $i < j$, the first transition occurs toward the neighboring cell with the lowest chemical potential, $\mu(1)$, etc. The transition probability from $0$ to $j$ is given by:
\begin{equation}
P_{0 \rightarrow j} \sim \Theta(\mu(0) - \mu_0) \times \textbf{Max}\left\{1, e^{-\beta (\mu(j) - \mu(0))} \right\}
\end{equation}
where $\Theta(x)$ is there to ensure that the donor cell is unstable.\\

The system undergoes two distinct regimes: In the first regime, the average chemical potential remains below the threshold value, i.e., $\langle \mu \rangle < \mu_0$, referred to as the SI regime, while in the second regime the average chemical potential saturates at the threshold, i.e., $\langle \mu \rangle = \mu_0$, marking the onset of the SII regime. In the SI regime, the average chemical potential $\langle \mu \rangle$ increases monotonically with the number of injected electrons $n$, until a critical point $n_0$ is reached. Beyond this point (SII), $\langle \mu \rangle$ stabilizes and remains approximately constant at $\mu_0$. This initial phase is inherently non-equilibrium in nature. In contrast, during the SII regime, the average chemical potential reaches a stationary statistical value, and the system gradually approaches thermodynamic equilibrium~\cite{najafi2019electronic}. If the time interval between successive electron injections is sufficiently long, allowing the system to reach a steady state, then the system is considered to have reached thermodynamic equilibrium. Conversely, if electrons are injected at a rate faster than the internal relaxation timescale, the system remains in a steady but non-equilibrium state~\cite{najafi2019electronic}.

\subsection{The quantities of interest}
The simulations are carried out on a triangular lattice of size $L\times L$ for various values of \( T \) and \( U \), while $Z_0$ is set to $ 0$, and $\mu_0 = 100$, \( \Delta = 1 \), and \( L = 128 \). The precise value of $Z_0$ is not critical for the simulations, as it can be effectively incorporated into the chemical potential. Equilibration is achieved after every 2000 electron injections, during which \( 100L^2 \) local equilibration steps occur. We then analyze the time series of the electronic avalanche sizes, \( S(t) \), where \( t \) represents the number of injections. Specifically, when the \( t \)-th avalanche occurs with a non-zero size, \( S(t) \) denotes the size of this avalanche.\\

Along with these quantities, we explore the dynamical properties of the model. To characterize the dynamical status of the model in terms of $T$ and $U$, we analyze the distibution of avalanche sizes $S$, which for scale free systems behave like
\begin{equation}
P(S)\propto S^{-\tau_s},
\label{Eq:stretched}
\end{equation}
where $\tau_s$ is the corresponding exponent that generally depends on $T$ and $U$.

\subsubsection{Autocorrelation function and power spectrum}
In systems exhibiting avalanche dynamics the time series of avalanche events often displays temporal correlations and scale-invariant behavior. The autocorrelation (AC) function in such systems captures the memory embedded in the dynamics: it reflects the memory inherent in a time series by measuring how the occurrence of one event affects the probability of future events occurring at later times. The auto-correlation function is defined as
 \begin{equation}     AC(t_0)\equiv\frac{\langle S(t)S(t+t_0)\rangle_t-\langle S(t)\rangle_t^2}{\langle S(t)^2\rangle_t-\langle S(t)\rangle_t^2},
 \label{Eq:ACF}
 \end{equation}
where the $t-$average of an arbitrary statistical observable is defined by $\langle O \rangle_t\equiv\frac{1}{t_{\text{max}}}\sum_{t=0}^{t_{\text{max}}}{O(S(t))}$.
Typically, a power-law decay of the autocorrelation function 
indicates long-range temporal correlations, a hallmark of self-organized criticality. The power spectrum is subsequently computed using the following relation:
 \begin{equation}
  PS(\omega)= \hspace{1mm} \lim\limits_{t_{\text{max}}\to\infty}\frac{1}{t_{\text{max}}}\left|{\int _{0}^{t_{max}}dt \hspace{1mm} S(t)\hspace{1mm}\exp\lbrack-i\omega t\rbrack }\right|^2,  
 \end{equation}
where $\omega$ is angular frequency, and $t_{\text{max}}$ is the maximum time in the analysis, and $|...|$ symbolizes the absolute value. For a stationary time series $\langle O(t)O(t^\prime) \rangle$ depends on $t-t^\prime$ (it is invariant under the transformation $t\to t+a$ and $t^\prime\to t^\prime+a$). For this case the power spectrum is a simple Fourier transformation of the autocorrelation function.  In self-similar time series, the power spectrum shows power-law behavior:
\begin{equation}
PS(\omega)\propto\frac{1}{\omega^{\alpha_{\text{PS}}}}
\label{Eq:PS}
\end{equation}
where $\alpha_{\text{PS}}$ is the exponent of power spectrum ~\cite{najafi2021flicker}. The exponents governing the autocorrelation decay, the avalanche size distribution, and the inter-event time distribution 
are not independent; they are often related through scaling laws derived from underlying stochastic or renormalization group arguments. For example, in many models, the autocorrelation exponent is linked to the avalanche duration exponent and the fractal dimension of event sequences, reflecting how the intrinsic temporal structure governs the statistical properties of avalanches. 
\section{resutls}
In this section, we present the results of our model. As stated in the previous section, in the stationary state (SII) the number of electrons entering the system from the reservoir is statistically equal to the number exiting ones through the boundaries. All of our investigation of the average density, along with other statistical quantities are performed in the SII regime.

\begin{figure}[H]
   \includegraphics[width=6cm, height=6cm]{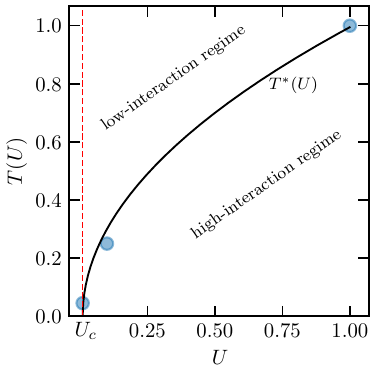} 
   \centering
    \caption{Schematic representation of the model's phase space. The diagram illustrates the qualitative behavior of the system across different regimes of the interaction parameter $U$. The boundaries are drawn to schematically highlight the transition regions, alongside of three real-data points: $(U,T)=$ $(0.01,0.045)$, $(0.1,0.25)$ and $(1.0,1.0)$ which represent the transition points obtained by simulations. We emphasis that the bold line is schematic, and has been designed to cross the real-data points. $U_c$ represents the point below which we have not observed high-interaction regime, i.e. only low-interaction exponents are observed.}
	\label{phase space}
   \end{figure}
Our results reveal that the system exhibits distinct behaviors in the large $U$ and small $U$ limits, characterized by distinct universal scaling exponents. This distinction is schematically represented in Fig.~\ref{phase space} (note that the separating line is illustrative and not derived from the actual data points). This conclusion is supported by analyses presented in the following sections, including the 
\begin{itemize}
    \item the scaling behavior of the average electron density with temperature,
    \item the power spectrum as a function of frequency,
    \item and the distribution and autocorrelation functions of electronic avalanche sizes.
\end{itemize}
\subsection{Average Electronic Density}
\begin{figure}[H]
\centering
   \includegraphics[width=8cm, height=6cm]{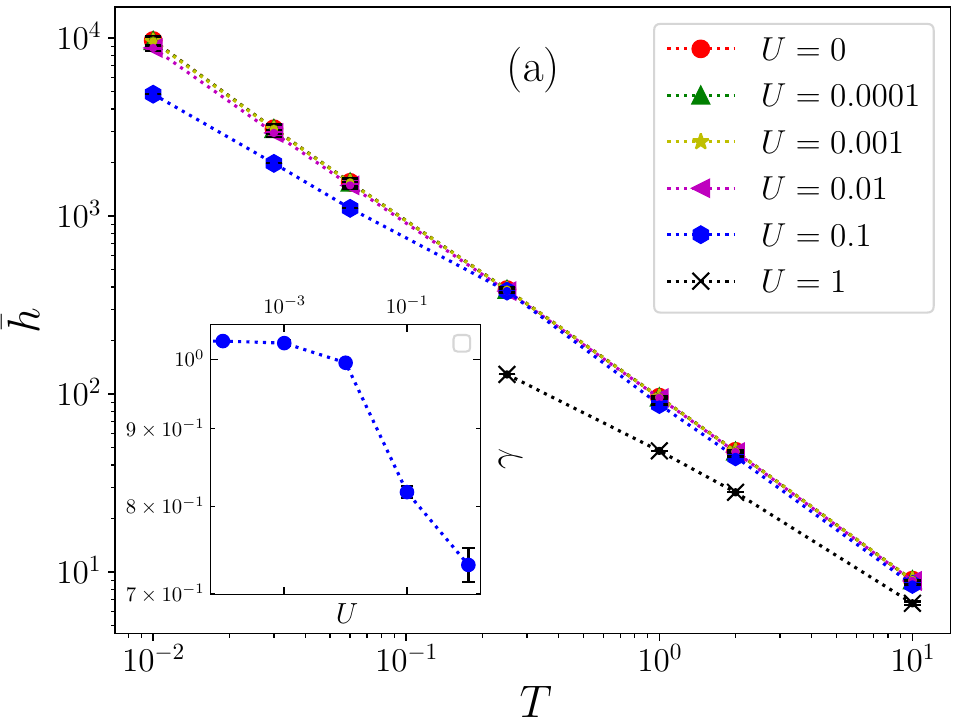}
   \includegraphics[width=8cm, height=6cm]{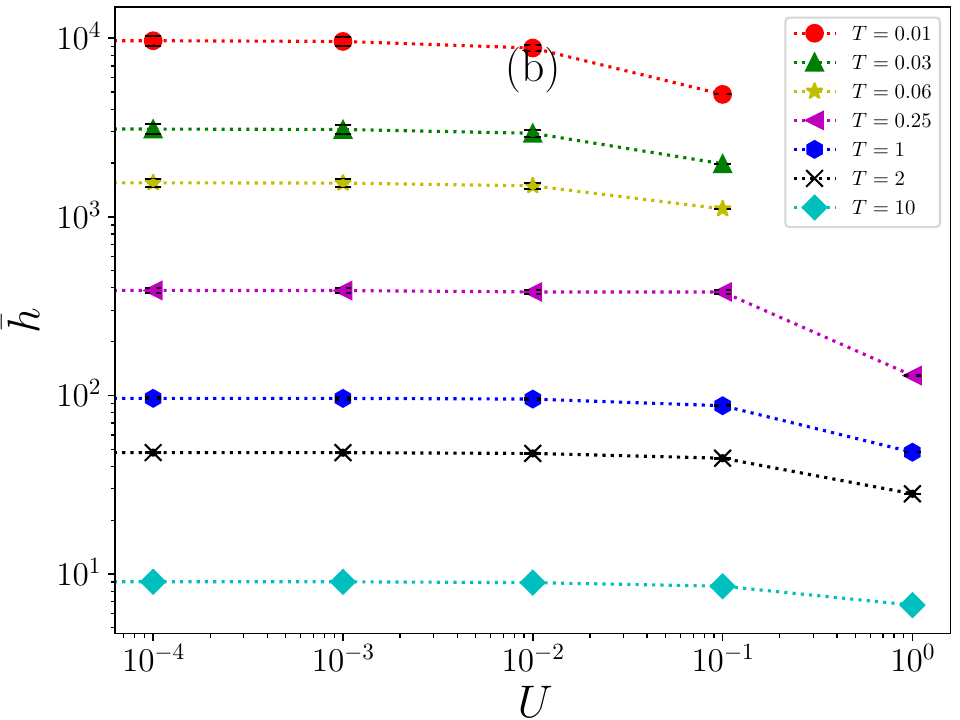}
   \caption{(a) Average electron density for in term of $T$ for various $U$ for $L=128$. (b) Average electron density for in term of $U$ for various $T$.}
\label{Fig:h}
\end{figure}
Figure~\ref{Fig:h} shows the behavior of the dimensionless electron density ($\bar{h}$), averaged over both space and ensemble, as a function of temperature for different values of $U$, as well as its variation with $U$ at several fixed temperatures. In this graph $\Delta =1 $ is held constant, and the system size is $L=128$. As seen in Fig.~\ref{Fig:h}a, for small $U$ values, $\bar{h}$ decreases with increasing temperature, exhibiting a power-law dependence: 
\begin{equation}
\bar{h}\propto T^{-\gamma},
\end{equation}
where the exponent $\gamma$ depends on $U$, presented in the inset. For very small $U$ values (low-interaction limit), it converges to a value that we call $\gamma_{\text{LIL}}$. For $U = 0.0001$ we found that $\gamma=1.02 \pm 0.02$. However, as $U$ increases, the exponent decreases—for instance, at $U = 1$ (high-interaction limit), it drops to $\gamma_{\text{HIL}} = 0.700 \pm 0.019$. An important observation here is that for intermediate $U$ values, there is a crossover point $T_{\text{CP}}$ above which $\gamma$ is consistent with the $\gamma_{\text{LIL}}$, and under which it is consistent with $\gamma_{\text{HIL}}$. $T_{\text{CP}}$ depends on $U$, the amounts of which are represented in the Fig.~\ref{phase space} (bold circles, note that $T_{\text{CP}}(U=0.1)\approx 0.2$). This reveals that the behaviour of the system at high and low $U$ values are different, showing different exponents. The dependence of $\bar{h}$ on $U$ is shown in Fig.~\ref{Fig:h}b where it is observed that $\bar{h}$ is nearly constant for small $U$ values, and begins to decrease significantly once $U$ exceeds a certain threshold.
\subsection{Power Spectrum of the Electronic Avalanche}
\begin{figure*}
   \includegraphics[width=8cm, height=6cm]{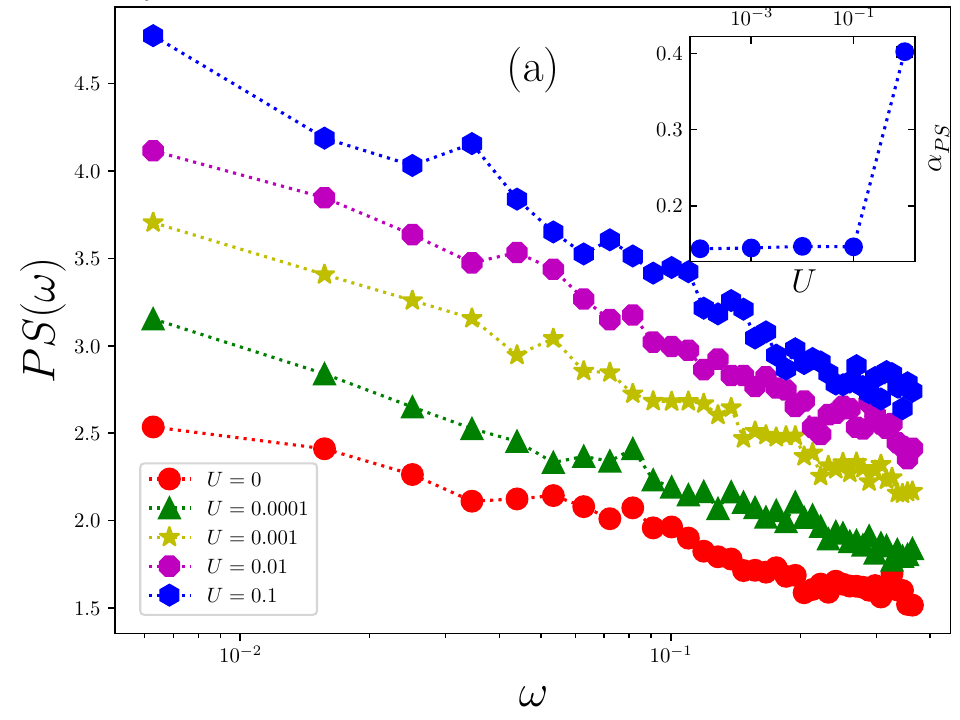} 
   \includegraphics[width=8cm, height=6cm]{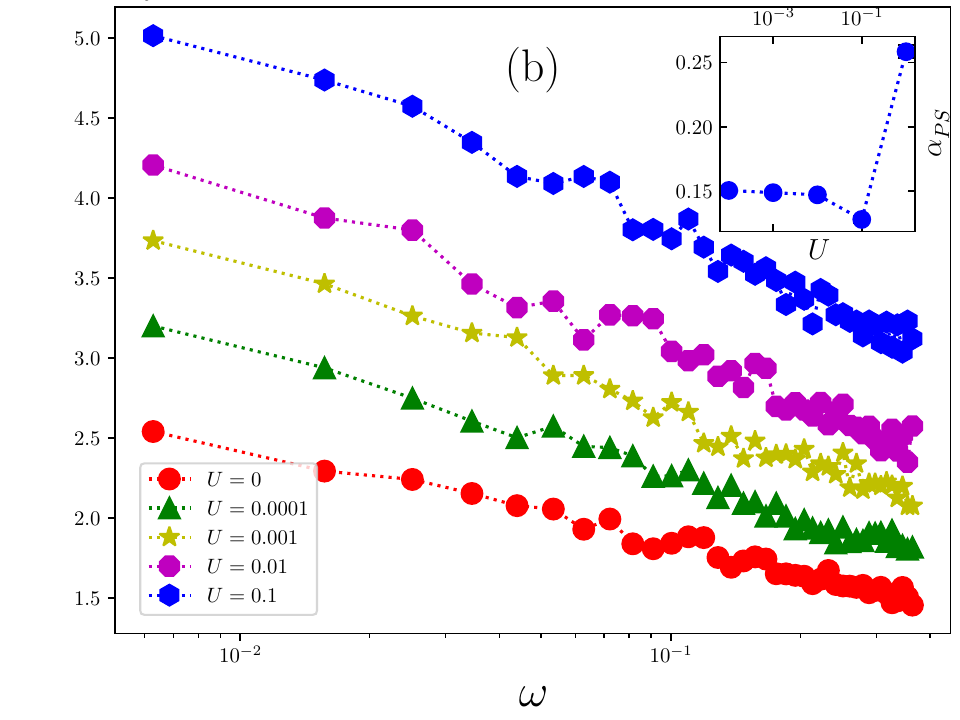}
   \includegraphics[width=8cm, height=6cm]{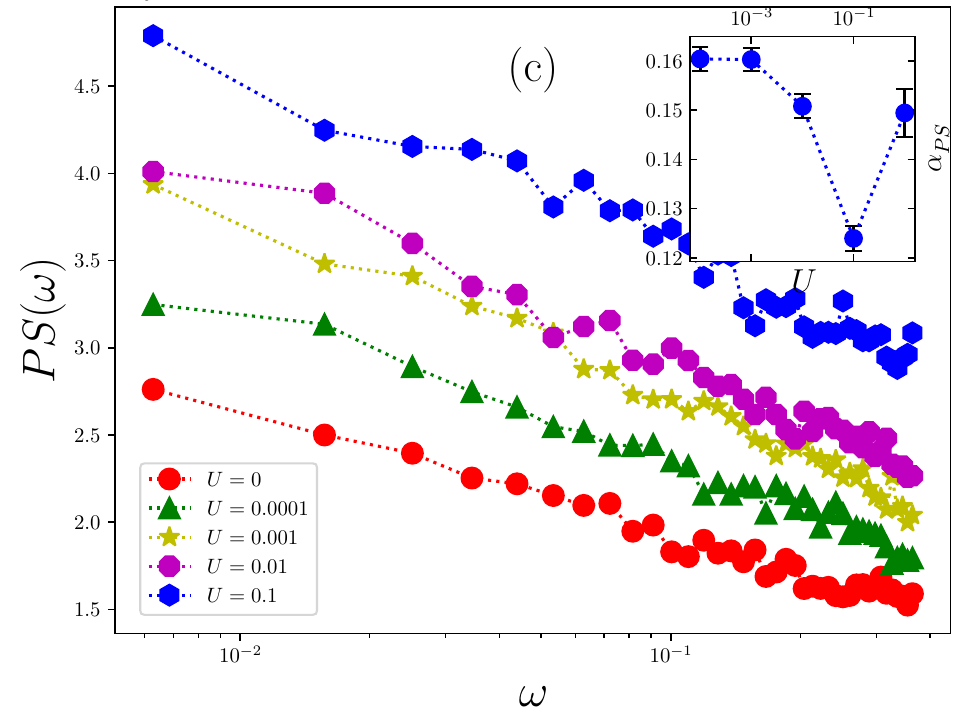} 
   \includegraphics[width=8cm, height=6cm]{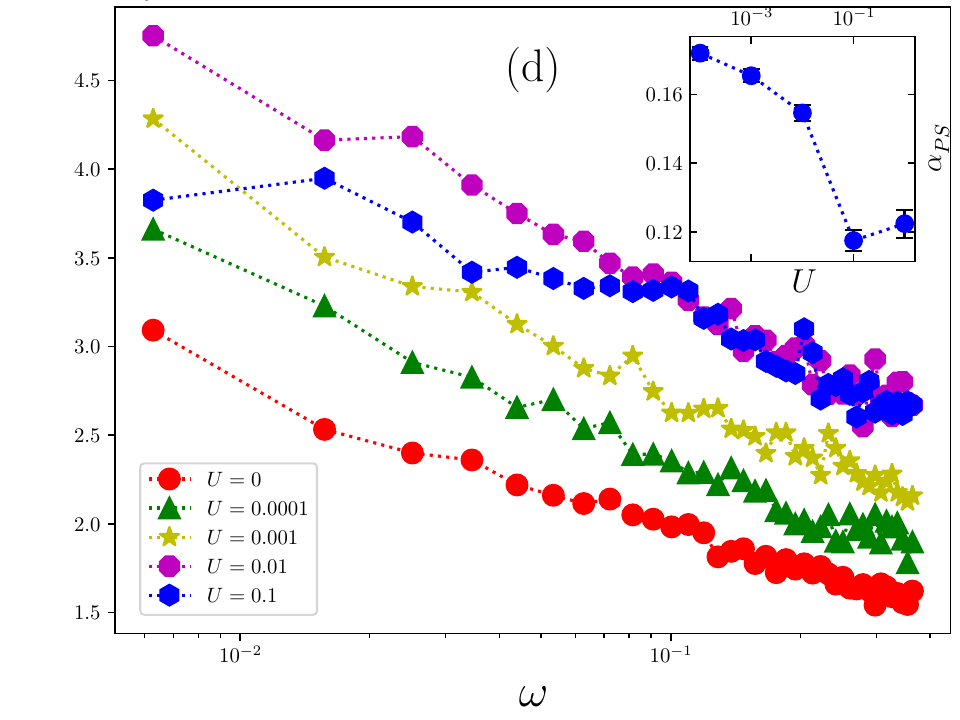}
   \caption{Power spectrum $PS(\omega)$ as a function of angular frequency $\omega$, calculated for system size 
$L=128$ and various values of interaction strength $U$, at different temperatures: (a) $T=0.25$, (b) $T=1.0$, (c) $T=2.0$, and (d) $T=10.0$.
(inset: $\alpha_{PS}$ in term of $U$ for various $T$.)}
	\label{powerspectrum}
\end{figure*}
\begin{figure}
\centering
   \includegraphics[width=7cm, height=4cm]{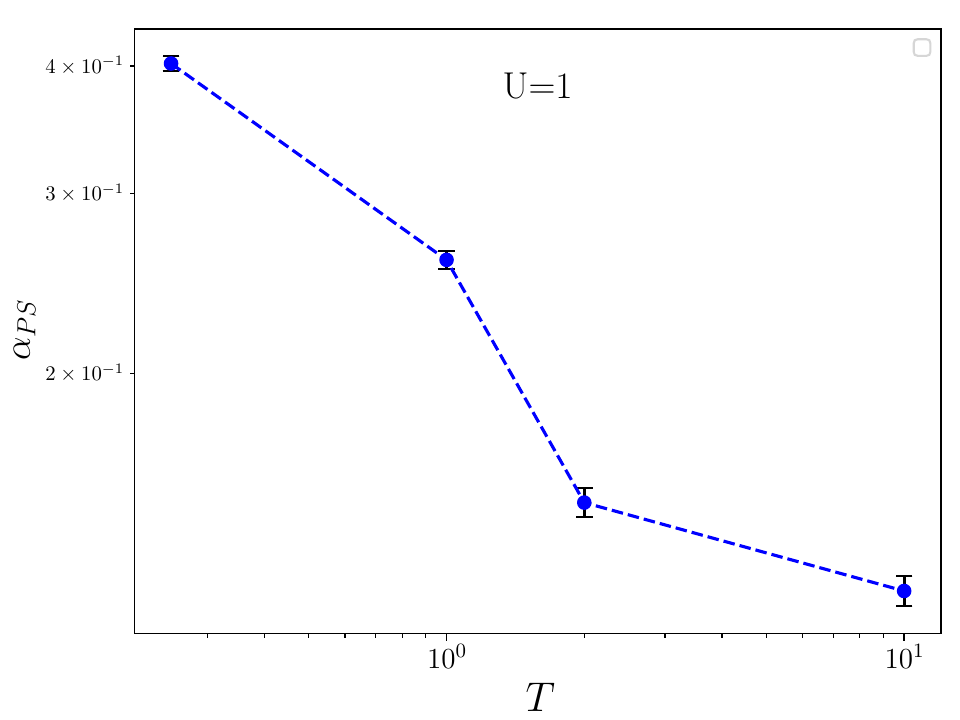} 
   \caption{Temperature dependence of $\alpha_{PS}$ under the condition $U = 1$.}
\label{Fig:alpha-LargeU}
\end{figure}

The power spectrum $PS(\omega)$ of the time series generated by electronic avalanches in terms of $\omega$ has been analyzed in Fig.~\ref{powerspectrum} for $L=128$ and $\Delta =1 $. Each graph, plotted for a fixed $T$ and various values of $U$, shows that the power spectrum behaves in a power-law fashion, with a well defined exponent $\alpha_{\text{PS}}$, typical for self-similar time series. More precisely, the equation~\ref{Eq:PS} applies here, where the exponent $\alpha_{\text{PS}}$ depends on both $U$ and $T$. In the insets, the variation of $\alpha_{\text{PS}}$ is shown in terms of $U$ for fixed temperatures $T$. For small temperatures $\alpha_{\text{PS}}$ shows a considerable jump at high $U$ values, e.g. for $T=0.25$ it jumps from $\alpha_{\text{PS}}^{(\text{LIL})}(U=0.1)=0.14\pm0.0024$ to $\alpha_{\text{PS}}^{(\text{HIL})}(U=1.0)=0.40\pm0.0066$. Based on this, we observe that, as $U$ increases, $\alpha_{\text{PS}}$ jumps from $\alpha_{\text{PS}}^{(\text{LIL})}$ (low interaction limit) to $\alpha_{\text{PS}}^{(\text{HIL})}$ (high interaction limit). While $\alpha_{\text{PS}}^{(\text{LIL})}(T)$ is nearly independent of $T$ for small $U$ values (where it is within the range $0.12<\alpha_{PS}<0.15$), $\alpha_{\text{PS}}^{(\text{HIL})}(T)$ depends considerably on $T$ in the large $U$ limit. Fig.~\ref{Fig:alpha-LargeU} shows $T$-dependence of $\alpha_{\text{PS}}$ for $U=1$. This figure indicates that $\alpha_{\text{PS}}$ drops almost one order of magnitude as $T$ varies from $T=0.25$ to $T=10$ in the large $U$ limit. This indicates a significant change in the underlying temporal correlations of the avalanche dynamics in the high interaction limit. Therefore, the large $U$ regime is characterized by sinsitivity of the noise exponent to the temperature. 
Noting that the noise exponent depends on (identifies) the Hurst exponent ($H$), we conclude that the pink noise statistics depends considerably on $T$ in the large $U$ limit. 
\subsection{Electronic Avalanche Autocorrelation and Distribution Functions}
\begin{figure}
   \includegraphics[width=4cm, height=4cm]{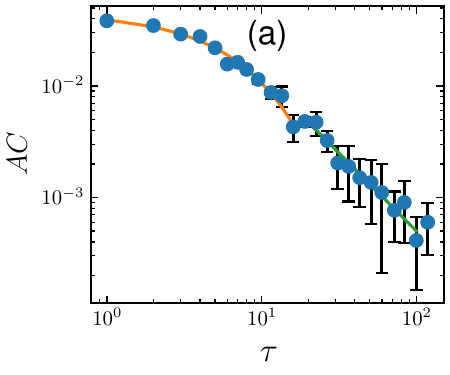}
    \includegraphics[width=4cm, height=4cm]{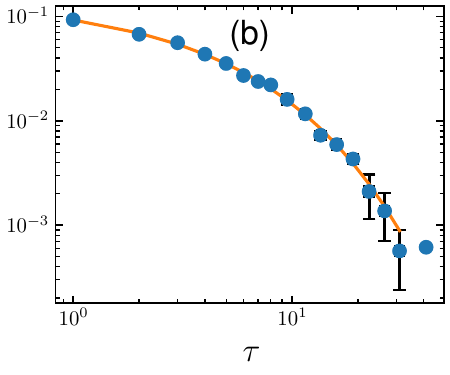}
   \caption{Autocorrelation function in term of $\tau$ for (a) $U=0.01,T=2$, and (b) $U=1,T=1$.}
	\label{AC}
\end{figure}

In this section we process the AC function based on the definition Eq.~\ref{Eq:ACF}. We calculate the exponents for a system of size $L = 128$ with a fixed value of $\Delta = 1$. Figure~\ref{AC}, represents the AC as a function of time for two different values $U=0.01$ and $U=1.0$. Note that AC is different for these two cases: it is mostly power-law for small $U$ limit, while it is different for large $U$ limit. More precisely, for $U=0.01$ (and other intermediate values of $U$) AC shows both behaviors: for small times it behaves like a \textit{stretched exponential form} (indicated by an orange line fit in Fig.~\ref{AC}a), while for large enough times, it behaves according to the power-law fashion (highlighted by a green line fit in Fig.~\ref{AC}a), as follows: 
\begin{equation}
\begin{split}
AC(t) &\sim e^{-a_1 t^{b_1}},\ \text{for } t<t^* \\
AC(t) &\sim t^{-b_2},\ \text{for } t>t^*
\end{split}
\label{Eq:piecewise}
\end{equation}
where $a_1$ and $b_1$ are positive constants characterizing the rate and nature of the decay, while $b_2$ is the exponent in the power-law, indicating the presence of long-range temporal correlations. The second equation applies to the power-law regime, which is observed at lower values of $U$. In this equation $t^*$ is a transition time between two regimes, which depends generally on $T$ and and $U$. The Hurst exponent is related to $b_2$ as follows for a pure power-law behavior~\cite{kantelhardt2002multifractal}:
\begin{equation}
H=\frac{2-b_2}{2}.
\label{Eq:Hurst}
\end{equation}
The piecewise behavior Eq.~\ref{Eq:piecewise} indicates a structural transition in the system's dynamics. In other words, increasing $U$ leads to a qualitative shift from power-law decay to exponential decay in the autocorrelation, reflecting a fundamental transformation in the underlying avalanche-like dynamics of the system.\\

\begin{figure*}
\centering
  \includegraphics{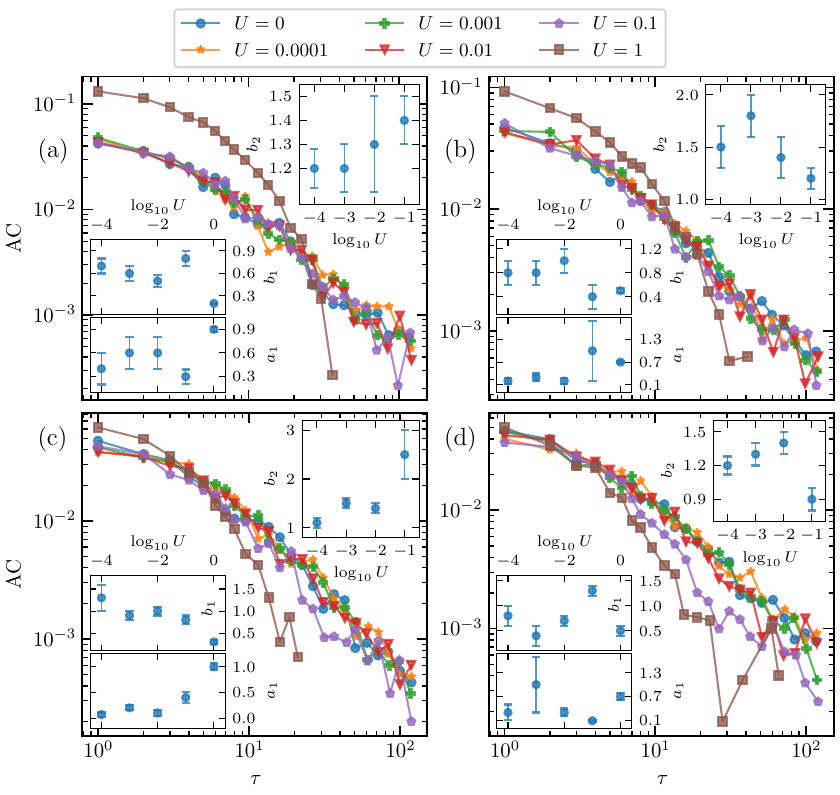}
   \caption{Autocorrelation function $AC(\tau) $ as a function of time delay $ \tau$, computed for system size $L=128$ and various values of interaction strength $U$, at different temperatures: (a) $T=0.25$, (b) $T=1.0$, (c) $T=2.0$, and (d) $T=10.0$.
   (inset: $\alpha_{AC}$ in terms of $U$ for various $T$ )}
	\label{autocorrelation}
\end{figure*}

In Fig.\ref{autocorrelation}, the  $AC(t)$ is analyzed for various values of the parameters $\beta$ and $U$, plotted as a function of time, exploring how the temporal correlations in the system evolve as a function of $U$ and $T$. $a_1$ (left-bottom intests), $b_1$ (left-top intests) and $b_2$ (right-top intests) are shown in each figure for (a) $T=0.25$, (b) $T=1.0$, (c) $T=2.0$ and (d) $T=10.0$, each for $U=0,0.0001,0.001,0.01,0.1,1.0$. These insets provide a comparative view of how the nature of the decay changes across the parameter space, highlighting the crossover between exponential and power-law regimes.\\

From the results presented in Fig.~\ref{autocorrelation}, for small $U$ values it is observed that the exponents remain more-or-less constant over a wide range of parameters. However, as $U$ becomes large, these exponents undergo a significant shift, indicating a structural change in the system’s temporal dynamics. For example, we see from Fig.~\ref{autocorrelation}a that $a_1$ exhibits a shift from $\approx 0.3$ at $U=0.1$ to $\approx 0.9$ at $U=1.0$ for $T=0.25$. A similar behavior is observed for the other exponents. This reinforces the idea that high values of $U$ lead to a transition from long-range correlated behavior to one dominated by short-range, rapidly decaying correlations.\\
\begin{figure*}
   \includegraphics[width=6cm, height=4cm]{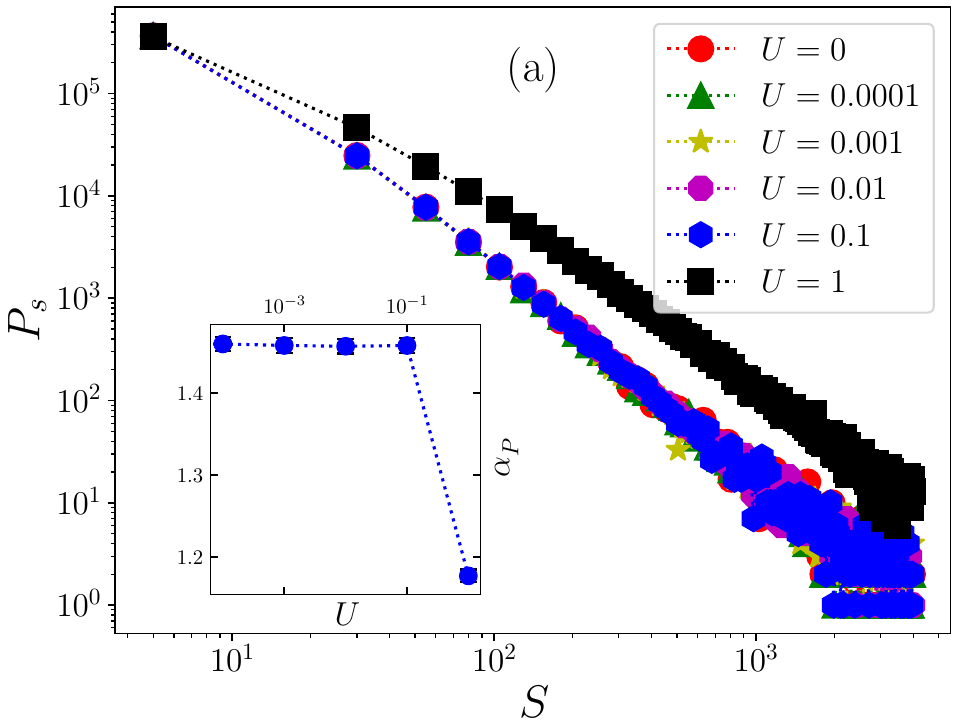}
    \includegraphics[width=6cm, height=4cm]{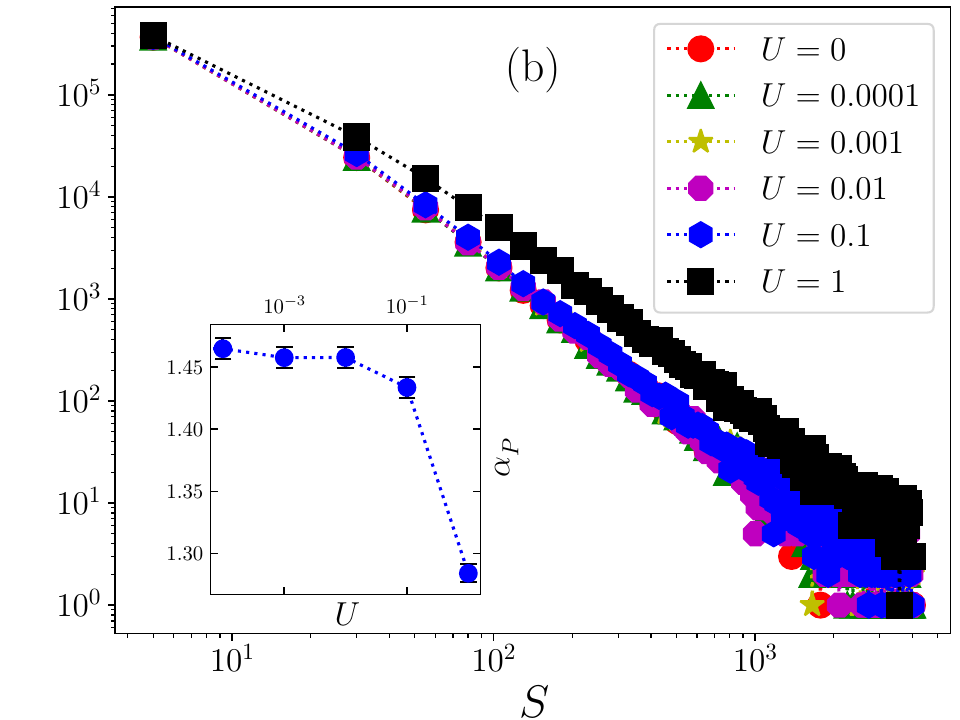} 
    \includegraphics[width=6cm, height=4cm]{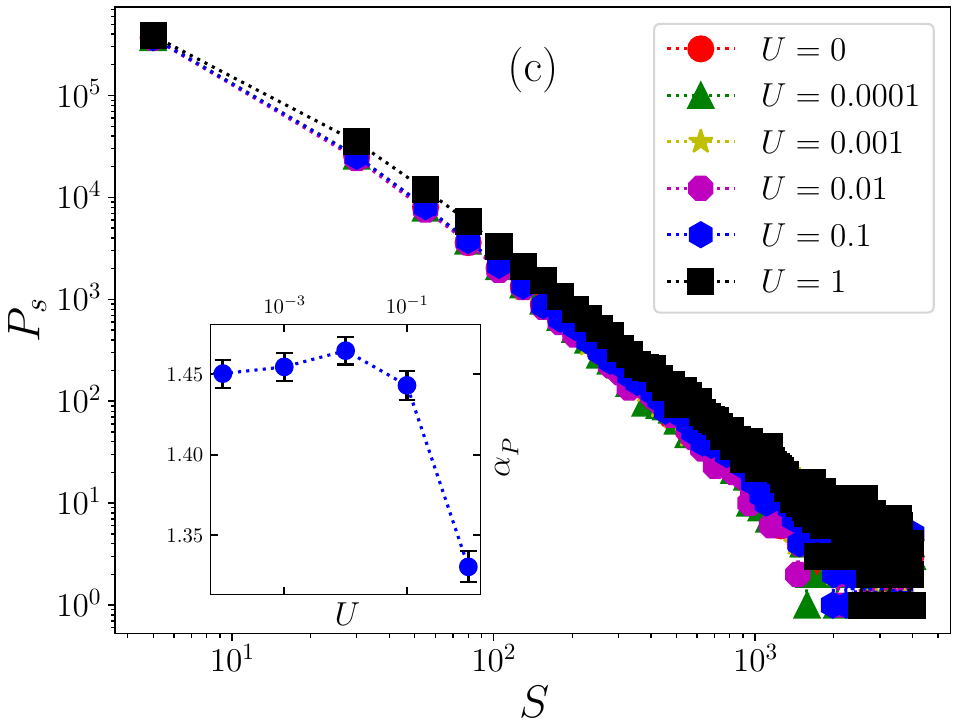} 
     \includegraphics[width=6cm, height=4cm]{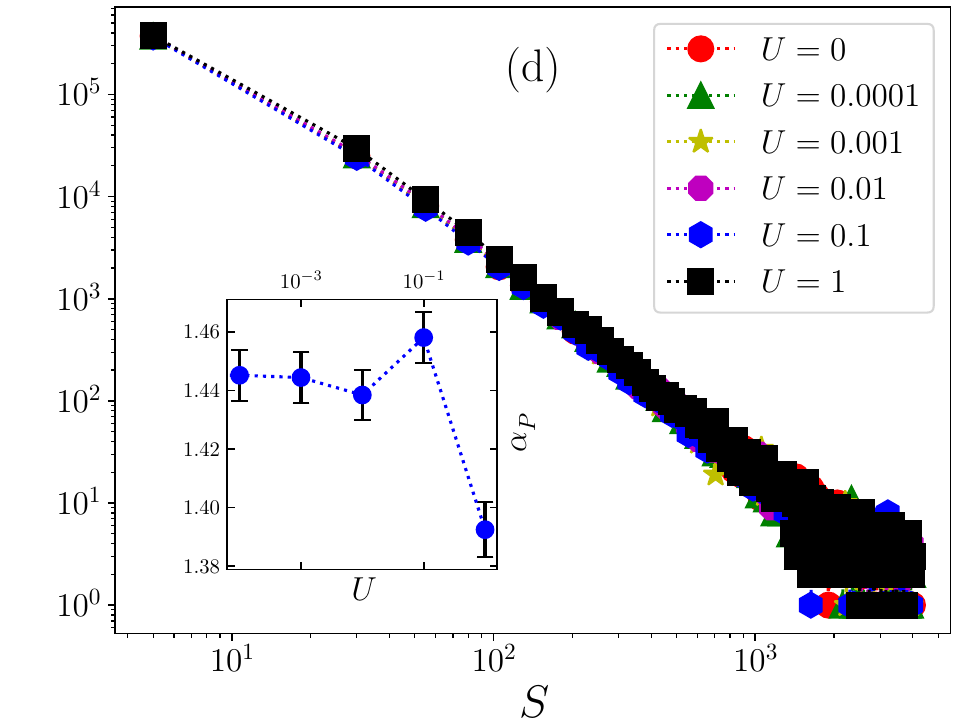} 
   \caption{Avalanche size distribution functions for different values of interaction strength $U$, computed at system size $L=128$ and at various temperatures: (a) $T=0.25$, (b) $T=1.0$, (c) $T=2.0$, and (d) $T=10.0$.
  ( inset: $\alpha_P$ in term of $U$ for various $T$ )}
	\label{distribution}
\end{figure*}

Another important quantity that reflects the state and underlying dynamics of the system is the distribution function of electronic avalanche sizes. In Fig.~\ref{distribution}, this distribution is plotted as a function of avalanche size $S$, for various values of $T$ and $U$, with the lattice size fixed at $L=128$ and $\Delta=1$. The plots illustrate how the probability of observing avalanches of different sizes depends on the system parameters, providing insight into the scaling behavior and possible criticality within the model. From the analysis of the distribution function graphs in Fig.~\ref{distribution}, it is observed that the distribution function of avalanche size shows a power-law fashion. Observe how the plot for $U=1$ is different from the others for low temperature values, indicating that in the large interaction limit the properties of the system becomes different. At higher temperatures, the effect of increasing $U$ on the avalanche size distribution becomes much less pronounced. This implies that thermal fluctuations dominate the dynamics at high temperatures, smoothing out the influence of $U$ and suppressing structural changes in avalanche behavior.\\

The insets of Fig.~\ref{distribution} present the exponent associated with the power-law behavior of the avalanche size distribution. From the analysis of this diagram, it is evident that at lower interaction strengths, the exponent remains nearly constant (around $1.45$ ) for all temperatures. However, at $U=1$, the exponent exhibits a noticeable drop to approximately a $T$-dependent value. This shift suggests that the system undergoes a change in its scaling behavior at higher $U$, even under low-temperature conditions. The reduction in the exponent indicates a relative increase in the likelihood of larger avalanches, reflecting a shift in the dynamics related to enhanced correlations and also structural changes within the system.

\section{Conclusion}
In this work, we have developed and numerically studied a self-organized critical model for a two-dimensional electron gas (2DEG) that explicitly incorporates electron-electron interactions. By discretizing the system based on coherence length and simulating avalanche dynamics, we have provided compelling evidence for the existence of two distinct universality classes governed by the strength of interactions.

Our results reveal a qualitative difference in critical behavior between the weakly and strongly interacting regimes. This distinction is substantiated by a comprehensive analysis of multiple observables: the scaling behavior of the average electron density with temperature (Fig.\ref{Fig:h}), the power spectrum of the electronic avalanche size time series (Fig.\ref{powerspectrum}), and the corresponding autocorrelation function (Fig.~\ref{autocorrelation}). Each of these observables displays a consistent shift in critical exponents as the interaction strength is varied, signaling a crossover between different scaling regimes.

These findings not only corroborate experimental observations of a qualitatively distinct metal-insulator transition in 2DEGs at low carrier densities (high interaction strength), but also establish a theoretical framework in which this transition can be understood in terms of self-organized criticality. The emergence of two universality classes—each characterized by its own set of scaling exponents—highlights the critical role of electron correlations in low-dimensional systems and opens avenues for further investigation into interaction-driven phase transitions beyond conventional paradigms.

\section*{Authors' Contributions}
M.N.N. conceived the research idea and supervised the project. The simulation code was initially written and developed by M.N.N., with further development by M.P., who also carried out the numerical calculations and data analysis. M.P. and V.A. prepared the figures. V.A. contributed to the interpretation of the data and results. M.P. drafted the initial version of the manuscript, and M.N.N. contributed to writing and refining the final version.

\bibliography{refs}
\end{document}